# Geometric Phases in Majorana Zero-Energy State

Zheng-Chuan Wang

Department of Physics & CAS Center for Excellence in Topological Quantum Computation, The University of Chinese Academy of Sciences, P. O. Box 4588, Beijing 100049, China.

## Abstract

The usual Berry phase for a Majorana zero-energy state is zero. In this manuscript, we propose a generalized geometric phase for Majorana zero-energy state, which is non-zero for the electron or hole, respectively. We calculate these non-zero geometric phases in a Ferromagnet (FI)/Topological Insulator (TI)/Superconductor (SC) hybrid system, whose magnetization can be manipulated by changing adiabatically the spin degree of freedom. The non-zero geometric phases have potential application on the topological quantum computation treatment of Majorana zero-energy modes. We also discuss the non-adiabatic geometric phase associated with Majorana zero-energy state by the path integral method.

## I. Introduction

In 1937, Majorana proposed a equation satisfied by a particle which is its own antiparticle[1]. Recently, this so-called Majorana Fermion has attracted more and more interests[2]. Besides looking for its realistic particle, attention has been paid on the condensed matter system to search for the quasiparticle excitations, i.e., a quasiparticle excitation in a superconductor is its own antiparticle due to the electron-hole symmetry, the zero-energy states (ZES) correspond to the Majorana bound states[3]. In 2008, Fu and Kane predicted theoretically that Majorana bound states can be found at the interface between topological insulators and superconductor[4]. Later researches indicated that Majorana bound states can even appear without topological insulator, such as in hybrid semiconductor with strong spin-orbital coupling in proximity to superconductors[5]. Nowadays, several experiments have demonstrated the evidences on the existence of Majorana bound states[6-9]. Following the suggestions by S. C. Zhang et al., chiral Majorana Fermion was also observed in a quantum anomalous Hall effect/superconductor hybrid device[10]. Since Majorana Fermion obeys non-abelian statistics, it has potential application on topological quantum computation[2].

Among treatments for topological quantum computation, the holonomic quantum computation scheme[11,12] proposed by Zanardi et al. in 1999 is a promising one, its fault-tolerance may be guaranteed by the topological character of geometric phase, the abelian and non-abelian geometric phase could realize all the geometric gates theoretically. Shortly after this, geometric quantum computation was achieved by Jones et al. in a nuclear magnetic resonance experiment[13], where the conditional Berry phase plays an important role. To meet the need of large-scale integrability, a Josephson superconductivity nanocircuit was applied to implement the gates for quantum computation, the interference effect of geometric phase was detected[14]. Till now, many systems have been utilized as candidates of geometric quantum computation, such as the trapped ions manipulated by laser[15], the semiconductor-based nanostructure driven by ultrafast laser pulse et al. [16]. Since the decoherence time in the practical system is very fast, the adiabatic geometric quantum computation cannot be completed during this short time, several authors developed geometric quantum computation by use of nonadiabatic conditional geometric phase[17-19], which make the geometric quantum computation as fast as possible.

If we exploit Majorana bound states to implement topological quantum computation, the dynamical phase will naturally disappear due to the zero-energy level, which avoids to adopt further techniques to get rid of the unnecessary dynamical phase. Unfortunately, the usual Berry phase for this Majorana zero-energy state (MZES) is also zero as pointed by Halperion et al. [20,21]. It seem that we cannot employ this Berry phase of MZES for topological quantum computation. However, things will be changed if we notice that the wavefunction of MZES is a spinor, the Berry phase contributed by the electron will cancel with the hole in this wavefunction, which leads to the zero Berry phase. It should be pointed out that the geometric phase of spinor wavefunction is somewhat different from the Berry phase for the usual scalar wavefunction. We ever investigated the geometric phase of spinor wavefunction in the relativistic Dirac equation and found that the usual Berry phase should be generalized[22]. In fact, the electron and hole in the spinor wavefunction will acquire their own geometric phases during an adiabatic procedure, which are non-zero. These non-zero geometric phases of MZES may have potential application on the topological quantum computation. In this manuscript, we will show how to obtain these non-zero geometric phases in the MZES.

## II. Geometric phases for MZES

Majorana Fermion excitation usually occurs at the interface of topological insulator or nanowire in proximity coupling to superconductors, which can be described by the Bogolyubov

de-Gennes Hamiltonian

$$\hat{H} = \begin{pmatrix} \hat{h} & \Delta_{sc} \\ \Delta_{sc}^{\dagger} & -\hat{h} \end{pmatrix},  \quad (1)$$

where $\hat{h}$ represents the single particle Hamiltonian which contains kinetic energy, potential, spin-orbital coupling and Zeeman term, $\Delta_{sc}$ is the superconductor pair potential. The system obeys the Bogolyubov de-Gennes equation: $\hat{H}\psi = E\psi$, in the Nambu notation, $\psi = ((\psi_{\uparrow}, \psi_{\downarrow}), (\psi_{\downarrow}^{\dagger}, -\psi_{\uparrow}^{\dagger}))$.

Suppose the system varies adiabatically with a slow parameter $R(t)$, then the time evolution of spinor wavefunction can be expressed as

$$\psi(R(t)) = \begin{pmatrix} \begin{pmatrix} u_{\uparrow}(R(t)) \\ u_{\downarrow}(R(t)) \end{pmatrix} \\ \begin{pmatrix} v_{\uparrow}(R(t)) \\ -v_{\downarrow}(R(t)) \end{pmatrix} \end{pmatrix} f(t), \quad (2)$$

where $v_{\uparrow}^{*}(R(t)) = u_{\uparrow}(R(t))$, $v_{\downarrow}^{*}(R(t)) = u_{\downarrow}(R(t))$, $f(t)$ is a time-dependent function. Substituting expression (2) into the following time-dependent BdG Schrödinger equation

$$i\hbar \frac{\partial}{\partial t}\psi(R(t)) = H(R(t))\psi(R(t)), \quad (3)$$

we will find $f(t)$ can be obtained as:

$$f(t) = e^{-\frac{i}{\hbar}\int_{0}^{t} E(R(t'))dt'} e^{\int_{0}^{R(t)}[\langle u_{\uparrow}(R(t'))|\nabla_R|u_{\uparrow}(R(t'))\rangle + \langle u_{\downarrow}(R(t'))|\nabla_R|u_{\downarrow}(R(t'))\rangle]dR'}$$
$$\cdot e^{\int_{0}^{R(t)}[\langle v_{\uparrow}(R(t'))|\nabla_R|v_{\uparrow}(R(t'))\rangle + \langle v_{\downarrow}(R(t'))|\nabla_R|v_{\downarrow}(R(t'))\rangle]dR'}, \quad (4)$$

which is just the usual dynamical phase and Berry phase associated

with Majorana Fermion, respectively. The Berry phase
$\int_0^{R(t')} [\langle u_\uparrow(R(t'))|\nabla_R|u_\uparrow(R(t'))\rangle + \langle u_\downarrow(R(t'))|\nabla_R|u_\downarrow(R(t'))\rangle + \langle v_\uparrow(R(t'))|\nabla_R|v_\uparrow(R(t'))\rangle + \langle v_\downarrow(R(t'))|\nabla_R|v_\downarrow(R(t'))\rangle] dR'$ is zero at all time because of $v_\uparrow^*(R(t)) = u_\uparrow(R(t))$ and $v_\downarrow^*(R(t)) = u_\downarrow(R(t))$ as stated by Halperin et al. [20,21].

However, we should emphasize that $u(R(t)) = \begin{pmatrix} u_\uparrow(R(t)) \\ u_\downarrow(R(t)) \end{pmatrix}$ corresponds to the wavefunction of electron, while $v(R(t)) = \begin{pmatrix} v_\uparrow(R(t)) \\ v_\downarrow(R(t)) \end{pmatrix}$ corresponds to the wavefunction of hole, they should have different time evolution, which means the spinor wavefunction (2) should evolve as follows:

$$\psi(R(t)) = \begin{pmatrix} \begin{pmatrix} u_\uparrow(R(t)) \\ u_\downarrow(R(t)) \end{pmatrix} f_1(t) \\ \begin{pmatrix} v_\uparrow(R(t)) \\ -v_\downarrow(R(t)) \end{pmatrix} f_2(t) \end{pmatrix}, \tag{5}$$

in which $f_1(t)$ and $f_2(t)$ are different functions, in the adiabatic evolution they can be determined by the time-dependent BdG Schrödinger equation

$$i\hbar \frac{\partial}{\partial t} \begin{pmatrix} \begin{pmatrix} u_\uparrow(R(t)) \\ u_\downarrow(R(t)) \end{pmatrix} f_1(t) \\ \begin{pmatrix} v_\uparrow(R(t)) \\ -v_\downarrow(R(t)) \end{pmatrix} f_2(t) \end{pmatrix} = E(R(t)) \begin{pmatrix} \begin{pmatrix} u_\uparrow(R(t)) \\ u_\downarrow(R(t)) \end{pmatrix} f_1(t) \\ \begin{pmatrix} v_\uparrow(R(t)) \\ -v_\downarrow(R(t)) \end{pmatrix} f_2(t) \end{pmatrix}, \tag{6}$$

where we have used the BdG equation $\hat{H}\psi = E\psi$ for an eigenstate. Solving Eq.(6), we have

$$f_1(t) = e^{-\frac{i}{\hbar}\int_0^t E(R(t'))dt'} e^{\int_0^{R(t)} \frac{\langle u(R(t'))| \nabla_{R'} | u(R(t'))\rangle}{\langle u(R(t'))| u(R(t'))\rangle} dR'} \tag{7}$$

and

$$f_2(t) = e^{-\frac{i}{\hbar}\int_0^t E(R(t'))dt'} e^{\int_0^{R(t)} \frac{\langle v(R(t'))| \nabla_{R'} | v(R(t'))\rangle}{\langle v(R(t'))| v(R(t'))\rangle} dR'} \tag{8}$$

The first terms in $f_1(t)$ and $f_2(t)$ stand for the dynamical phases which are zero for the MZES, while the imaginary part of second terms $\gamma_u = \text{Im}\int_0^{R(t)} \frac{\langle u(R(t'))| \nabla_{R'} | u(R(t'))\rangle}{\langle u(R(t'))| u(R(t'))\rangle} dR'$ and $\gamma_v = \text{Im}\int_0^{R(t)} \frac{\langle v(R(t'))| \nabla_{R'} | v(R(t'))\rangle}{\langle v(R(t'))| v(R(t'))\rangle} dR'$ are just the geometric phases we want, their real parts describe the usual time evolutions of $u(R(t))$ and $v(R(t))$ except the dynamical and geometric phases. We can see that the geometric phase $\gamma_u = -\gamma_v$, because

$$\int_0^{R(t)} \frac{\langle u(R(t'))| \nabla_{R'} | u(R(t'))\rangle}{\langle u(R(t'))| u(R(t'))\rangle} dR' = \int_0^{R(t)} \left(\frac{\langle v(R(t'))| \nabla_{R'} | v(R(t'))\rangle}{\langle v(R(t'))| v(R(t))\rangle}\right)^* dR'.$$

So far, we have obtained the non-zero geometric phases for the MZES, the electron and hole have their own geometric phases $\gamma_u$ and $\gamma_v$, respectively, they are the generalization of usual Berry phase in spinor wavefunction.

If we denote $\vec{A}_u = \frac{\langle u(R(t))| \nabla_R | u(R(t))\rangle}{\langle u(R(t))| u(R(t))\rangle}$, we can prove that $\vec{A}_u$ is a gauge potential, because when we make a gauge transformation $|u'\rangle = e^{i\Theta(R)}|u\rangle$, where $\Theta(R)$ is an arbitrary real function, $\vec{A}_u$ will

transform as:

$$\vec{A}'_u = \frac{\langle u'(R(t))| \nabla_R | u'(R(t))\rangle}{\langle u'(R(t))| u'(R(t))\rangle} = \frac{\langle u(R(t))| \nabla_R | u(R(t))\rangle}{\langle u(R(t))| u(R(t))\rangle} + \nabla\Theta(R) \quad (9)$$
$$= \vec{A}_u + \nabla\Theta(R)$$

which is exactly the gauge transformation for $\vec{A}_u$. For a closed loop $C$ in $R$ space, the geometric phase $\gamma_u(C) = \oint_C \vec{A}_u \cdot d\vec{R}$. By use of Stokes theorem, we have

$$\gamma_u(C) = \iint_S \vec{B}_u \cdot d\vec{S}, \quad (10)$$

where $\vec{B}_u = \vec{\nabla} \times \vec{A}_u$ is the curvature corresponding to gauge potential $\vec{A}_u$, $S$ is the surface enclosed by loop $C$. Eq.(9) and (10) demonstrate the topological characters of geometric phase $\gamma_u$. Similarly, $\vec{A}_v = \frac{\langle v(R(t))| \nabla_R | v(R(t))\rangle}{\langle v(R(t))| v(R(t))\rangle}$ is also the gauge potential, we can obtain its curvature as $\vec{B}_v = \vec{\nabla} \times \vec{A}_v$, too.

### III. Geometric phases in FI/TI/SC junction

Consider a Ferromagnet/Superconductor junction on a strong topological insulator surface as shown in Fig.1. The bulk s wave supercondcutor interacts with the edge state electrons of TI by the proximity effect and induces the superconductivity in the topological surface states, the ferromagnet in the edge state is also induced by the proximity effect of the ferromagnet insulator (FI), the Hamiltonian for this system reads:

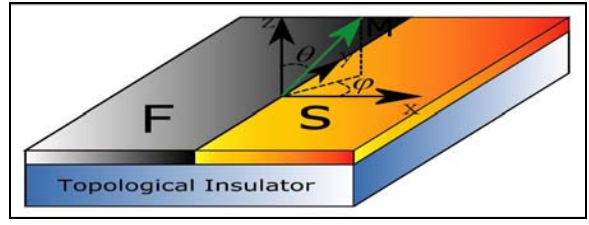

Fig.1 Schematic structure of a Ferromagnet/Superconductor junction on the surface of a topological insulator, where $\theta$ and $\alpha$ are the polar angle and azimuthal angle for magnetization.

$$\hat{H} = \begin{pmatrix} v_f \vec{\sigma} \cdot \vec{p} - \mu + \vec{m}(r) \cdot \sigma & \Delta_{sc}(r) \\ \Delta_{sc}^{\dagger}(r) & -v_f \vec{\sigma} \cdot \vec{p} + \mu + \vec{m}(r) \cdot \sigma \end{pmatrix}, \quad (11)$$

where $v_f$ is the Fermi velocity of edge states, $\mu$ is the chemical potential. $\vec{\sigma} = (\sigma_x, \sigma_y, \sigma_z)$ is the Pauli matrix. $\vec{m}(x)$ is the piecewise magnetization of FI, which can be chosen as a soft magnetic insulator and controlled by a weak external magnetic field. Luo et al. ever proposed a scheme to steer the spin degree of freedom of MZES by a varying magnetization $\vec{m}(x)$ for the topological quantum computation [21], so we can adopt $\vec{m}(x)$ as a slow parameter of the adiabatic procedure.

For simplicity, we only consider the one-dimensional transport along x-axis, the MZES in the FI and SC region can be written as[21]:

$\psi_{FI}(x) = a_e(e^{-ik_m x}, e^{i(\alpha+\varphi)}e^{-ik_m x}, 0, 0) +$
$a_e^*(0, 0, e^{-i(\alpha+\varphi)}e^{ik_m x}, e^{-ik_m x})^T e^{k_{FI} x}(x < 0)$

and

$\psi_{SC}(x) = m_e(ie^{ik_{SC} x}, 0, e^{ik_{SC} x}, 0) +$
$m_e^*(0, e^{-ik_{SC} x}, 0, ie^{-ik_{SC} x})^T e^{K_{SC} x}(x > 0)$,

respectively, where the parameters $\varphi$, $k_{FI}$, $k_m$, $k_{SC}$, $K_{SC}$ are defined as:

$$e^{i\varphi} = \frac{i\sqrt{m_\parallel^2 - \mu_{FI}^2} + \mu_{FI}}{m_\parallel}, \quad k_{FI} = \frac{\sqrt{m_\parallel^2 - \mu_{FI}^2}}{\hbar v_f},$$

$$K_{SC} = \frac{\Delta}{\hbar v_f}, \quad k_{m\,SC} = \frac{m_\parallel / \mu_{SC}}{\hbar v_f}$$

with $m_\parallel = |m|\sin\theta$ is the in-plane component of magnetization, $\theta$ is the polar angle, $\alpha$ is the azimuthal angle for magnetization. The coefficients $a_e$, $a_e^*$, $m_e$, $m_e^*$ may be determined by the matching condition at $x = 0$: $\psi_{FI}(0) = \psi_{SC}(0)$. If we fix the polar angle $\theta$ and change the azimuthal angle $\alpha$ from $0 \sim 2\pi$, according to Eq.(7) and (8), the geometric phases for the electron and hole in the FI region is $\gamma_{u(FI)} = \pi$ and $\gamma_{v(FI)} = -\pi$, respectively, while the geometric phases for the electron and hole in SC region are: $\gamma_{u(SC)} = 0$ and $\gamma_{v(SC)} = 0$. The geometric phases in MZES have potential application on topological computation.

## IV. Summary and Discussions

We present the non-zero geometric phases for MZES, the electron and hole may acquire their own geometric phases during an adiabatic procedure, that is different from the usual Berry phase which is zero for MZES. We expect that these non-zero geometric phases may have potential application on the topological quantum computation.

However, we only investigate the adiabatic geometric phases for MZES in the above, there exists nonadiabatic geometric phase during a nonadiabatic procedure, which is convenient for us to implement the rapid topological quantum computation, so it is worthwhile to explore.

For the time evolution of a system with period $T$, we have

$$|\psi(r,T)\rangle = T\exp(-i\int_0^t \frac{\hat{H}(R(t'))}{\hbar}dt')|\psi(r,0)\rangle, \qquad (15)$$

where $T$ denotes the time order operator. After a period $T$, we have

$$|\psi(r,T)\rangle = \exp(i\phi)|\psi(r,0)\rangle. \qquad (16)$$

Remembering that the dynamical phase for MZES is zero, $\phi$ is just the nonadiabatic geometric phase we want, it can be expressed as:

$$e^{i\phi} = \langle\psi(r,0)|T\exp(-i\int_0^t \frac{\hat{H}(R(t'))}{\hbar}dt')|\psi(r,0)\rangle. \qquad (17)$$

The calculation of nonadiabatic geometric phase $\phi$ is complicated, we will study it by the path integral method. After make a time-discretization $t_0, t_1, \ldots t_N$ as done by Kuratsuji et al.[23], where $t_i - t_{i-1} = \varepsilon = \frac{T}{N}$, we have

$$e^{i\phi} = \int T_N(C)\exp(\frac{i}{\hbar}S_0(C))\prod d\mu(r_t, p_t), \qquad (18)$$

with $S_0(C) = \int(p\dot{r} - H_0)dt$ and

$$T_N(C) = \sum_{m_1}\ldots\sum_{m_{N-1}}\langle\psi(r,0)|\exp(-ih(N)\varepsilon/\hbar)|m_{N-1}\rangle\ldots \\ \langle m_k|\exp(-ih(k)\varepsilon/\hbar)|m_{k-1}\rangle\ldots\langle m_1|\exp(-ih(1)\varepsilon/\hbar)|\psi(r,0)\rangle \qquad (19)$$

where the Hamiltonian $\hat{H}$ has been divided as $\hat{H} = H_0 + h$, $H_0$ is independent of the parameter $R(t)$, while $h$ changes with $R(t)$, $\{|m_k\rangle\}$ are the complete eigenstates of $h(R)$.

As pointed out by Kuratsuji et al.[23], the quantum transitions only occur between states with the same quantum number $m_k$ during an adiabatic procedure, i.e., $\langle m_k | \exp(-i h(k)\varepsilon / \hbar)|m_k\rangle$, it will contribute an adiabatic Berry phase and an dynamical phase[23], which are both zero for MZES. In the next, we will continue to explore the nonadiabatic geometric phase that hadn't been studied by Kuratsuji et al.. In the nonadiabatic case, the quantum transitions will occur between the states with different quantum number, i.e.

$$\langle n_k | \exp(-i h(k)\varepsilon / \hbar) | m_{k-1}\rangle = \exp(-i E_k \varepsilon / \hbar)\langle n_k(R_k) | m_{k-1}(R_{k-1})\rangle. \quad (20)$$

For the MZES, the dynamical phase $-i E_k \varepsilon / \hbar$ is zero, in the limit of $\varepsilon \to 0$

$$\langle n_k(R_k) | m_{k-1}(R_{k-1})\rangle \simeq 1 - \langle n_k | \frac{\partial}{\partial R_k} | m_{k-1}\rangle \Delta R_k \simeq \exp(i \omega_{n_k m_{k-1}}), \quad (21)$$

where $\omega_{n_k m_{k-1}} = \langle n_k | \frac{\partial}{\partial R_k} | m_{k-1}\rangle \Delta R_k$.

So the product of all the quantum transitions $\{\langle n_k | \exp(-i h(k)\varepsilon / \hbar) | m_{k-1}\rangle\}$ in $T_N(C)$ will contribute a nonadiabatic geometric phase:

$$T_N(C) = \exp(i \phi) = \exp(i \omega_{m_N m_{N-1}})\ldots \exp(i \omega_{m_k m_{k-1}})\ldots \exp(i \omega_{m_1 m_0}). \quad (22)$$

Certainly, the practical calculation of nonadiabatic geometric phase

$\phi$ is very complicated, an alternative way is the path integral quantum Monte Carlo method, which is powerful for the path integral calculation, we will left it for further explore.

## Acknowledgments

This study is supported by the National Key R&D Program of China (Grant No. 2018FYA0305804), and the Key Research Program of the Chinese Academy of Sciences (Grant No. XDPB08-3).